\documentstyle[12pt,prl,aps]{revtex} 
\def\addcontentsline#1#2#3{\relax}

\begin{document}
\author{Yshai Avishai$^{\,1,2}$, Daniel Berend$^{\,2}$, 
and Richard Berkovits$^{\,3}$}
\address{${}^{1}$Department of Physics, Ben-Gurion University 
of the Negev, Beer-Sheva 84105, Israel \\
${}^{2}$Department of Mathematics and Computer Sciences,
Ben-Gurion University of the Negev, Beer-Sheva 84105, Israel \\
${}^{3}$Minerva Center, Department of Physics, Bar Ilan University,
Ramat Gan 52900, Israel}  
\title{Fluctuation of Conductance Peak Spacings 
in Large Semiconductor Quantum Dots}

\maketitle
\bigskip

\begin{abstract}
Fluctuation of Coulomb blockade peak spacings 
in large two-dimensional semiconductor quantum dots are studied 
within a model based on the electrostatics 
of several electron islands among which there are
random inductive and capacitive couplings.
Each island can accommodate 
electrons on quantum orbitals whose 
energies depend also on an external magnetic field. 
In contrast with a single island 
quantum dot, where the spacing distribution is 
close to Gaussian,
here the distribution has a peak at small 
spacing value. The fluctuations are mainly due 
 to charging effects. 
 The model 
can explain the occasional occurrence of 
couples or even triples of closely spaced Coulomb blockade peaks, 
as well as the qualitative behavior of peak positions 
with the applied magnetic field. 
 \\

\end{abstract}

\pacs{PACS number: }
%72.10.-d, 72.10.Bg, 73.40.Gk}

\newpage
\section {Introduction}
Recently, it became apparent that 
the physics exposed in the addition spectra of quantum dots is 
rather rich, and hence its investigation is
at the focus of both experimental and theoretical studies. 
The present work concentrates on the distribution of spacings 
between Coulomb blockade peaks in large semiconductor quantum dots.
Coulomb blockade is evidently one of the hallmarks of 
mesoscopic physics. The experimental achievement of tracing an 
addition of a single electron to a quantum dot and the appearance 
of isolated conductance peaks led to the concept of 
single electron transistors. After 
the origin of Coulomb blockade peaks has been elucidated, investigation 
is directed toward more subtle questions like their heights, widths 
and spacings. The underlying physics is related 
to the ground state energy, chemical potential 
and inverse compressibility of a few electron islands
coupled capacitively to its environment, 
as well as fluctuations of these quantities
with the number $N$ of electrons on the dot. 

\noindent
As far as the distribution of spacings between adjacent
Coulomb blockade peaks is concerned, the 
question can be stated as follows: 
According to the simplest picture in which the 
quantum dot is regarded as a single electron island 
whose coupling with the leads is through 
its capacitance $C$, 
the total potential energy of a 
quantum dot is $Q^2/2C-V_g Q$, where $V_g$ is 
the corresponding gate voltage. The conductance peaks 
occur at those values of $V_g$ for which $C V_{g}=e (N+1/2)$,
where $e$ is the electron charge (henceforth $e=-1$).
For this value of $V_g$ the
addition of an electron to the dot (which contains 
$N$ electrons) does not cost any 
charging energy.
The position of the $N^{th}$ Coulomb blockade peak is then a linear 
function of $N$, and therefore
the spacing should be a constant $1/C$, independent 
of $N$. Recent experiments \cite{Sivan,Kouwenhoven} indicate 
however that spacing between Coulomb blockade peaks in small 
quantum dots is in general not constant but, rather, a fluctuating 
quantity close to Gaussian. The average of its distribution approximately 
coincides with the constant value mentioned above,
but the elucidation of its standard deviation is 
still under investigation \cite{Mirlin,Orgad}. 

\noindent
The situation is even more 
intriguing if the quantum dot is very large. 
As indicated in a series of recent experiments, 
the spacing occasionally vanishes, 
namely, two peaks (and sometimes even three peaks) coincide. Moreover, 
the evolution of peak positions and spacings with an applied magnetic 
field indicates the existence of strong correlations between 
them \cite{Ashoori1,Ashoori2,Ashoori3}. 
These observations motivated numerous theoretical 
models based on the concept of pair tunneling \cite{Wan} 
or that of two-electron bound-states in depleted 
electron islands \cite{Raikh}.\\
\noindent
In the present work we examine the scenario 
according to which a large quantum dot,
like the one used in the last experiment \cite{Ashoori3}, is, 
in fact, composed of {\em several} electron islands which are 
coupled capacitively among themselves as well as 
to the leads. Electrons are added 
in such a way that the total potential 
energy of the dot is minimum. This simple generalization of 
the single island picture leads to a remarkable change in the 
spacing distribution from a Gaussian \cite{Sivan,Mirlin,Orgad} centered 
around a finite average to 
one which is large at small spacings. When the coupling between 
islands is zero, the distribution has indeed a maximum at 
zero spacing. If the coupling is non-zero but small, the 
maximum is slightly shifted away from zero, yet leaving 
an appreciable tail down to zero. 
These results are short of explaining the perfect 
overlap of peaks, since it requires a delta function component
at zero spacing. Yet, it leads to the occurrence of couples and 
sometimes triples of closely spaced peaks, similar to the experimental
observation. Moreover, the evolution of the peak positions 
with the magnetic field is qualitatively similar to the 
experimental one. On the other hand, the present model does 
not predict a definite periodicity in the bunching of Coulomb 
blockade peaks with electron number $N$. 
In the next section the model is explained, and 
the results of the calculations are presented in section 3.
\section{Formalism}
We consider a large isolated two-dimensional quantum dot 
in a perpendicular magnetic field $B$ subject to a gate 
voltage $V_{g}$. Unlike the traditional 
Coulomb blockade picture, it might contain {\em several} 
electron islands which can be regarded as metallic 
objects with inductive couplings among themselves. 
These are determined by a positive 
definite symmetric matrix $C$ whose
diagonal elements $C_{ii} \equiv C_{i}$ -- the 
corresponding capacities -- are positive, whereas the 
nondiagonal elements $C_{ij}=C_{ji}, (i \ne j)$ -- the corresponding 
coefficients of induction -- are negative.
The electrostatic energy of such a system may be written as
\begin{eqnarray}
E_{c}= \frac {1} {2} \sum_{i,j=1}^{K} p_{ij} N_{i} N_{j}- V_{g} N,
\label{eq_Ec}
\end{eqnarray}
where $N_{i}$ is the number 
of electrons on island $i$ 
(the number of islands $K>1$ might be around $10$), 
$N=\sum_{i=1}^{K} N_{i}$ is the 
total numbers of electrons, and $p=(p_{ij})$ is the (symmetric positive-definite)
matrix $C^{-1}$. 

\noindent 
Beside the electrostatic energy it is assumed that electrons in each island
occupy single particle quantum states (orbitals),
whose energies $\epsilon_{i \alpha}$ ($i=1,2,..K; \alpha=1,2,..$) 
depend on the confining potential as well as on the
magnetic field. The latter is manifested
through its orbital effects as well 
as due to Zeeman splitting (in which case 
the quantum number $\alpha$ contains also a spin label). 
The corresponding occupation numbers 
$n_{i \alpha}$ can be either $0$ or $1$. 
The system described above might then be represented by
a classical Hamiltonian 
\begin{eqnarray}
H=H_{c} + H_{sp}, 
\label{eq_H}
\end{eqnarray}
where the charging Hamiltonian $H_{c}$ is just the electrostatic 
energy of (\ref{eq_Ec}) written in terms of the orbital 
occupation numbers
\begin{eqnarray}
H_{c}= \frac {1} {2} \sum_{i,j=1}^{K} p_{ij} 
[ \sum_{\alpha} n_{i \alpha} ]
[ \sum_{\alpha '} n_{j \alpha '} ]
-V_{g} \sum_{i=1}^{K} \sum_{\alpha} n_{i \alpha},
\label{eq_Hc}
\end{eqnarray}
and the single particle part of the Hamiltonian $H_{sp}$ is
\begin{eqnarray}
H_{sp}=\sum_{i=1}^{K} \sum_{\alpha} \epsilon_{i \alpha} n_{i \alpha} .
\label{eq_Hsp}
\end{eqnarray}
The precise form of the matrix elements $p_{ij}=[C^{-1}]_{ij}$
as well as the single particle energies $\epsilon_{i \alpha}$ 
are specified in the next section when we present our results.
Despite the fact that the Hamiltonian $H=H_{c} + H_{sp}$ is classical 
(and relatively simple), the elucidation of its spectrum
for large $N$ and $K$ is virtually hopeless. In order 
to compute the ground state energy $E(N)$,
one has to find the minimum of $H$ 
on all the possible $K$-tuples $(N_{1},N_{2},\ldots,N_{K})$ with the 
constraint $\sum_{i=1}^{K} N_{i}=N$.  Note that 
the so-called ``Coulomb Glass'' model obtains 
as a special case when $i$ refers 
to a lattice site with random energy 
$\epsilon_{i \alpha}=\epsilon_{i}$,
and a single orbital $N_{i}=n_{i}=0,1$. The interaction matrix is
then given by $p_{ij}=1/r_{ij}$ for $i \ne j$ and $p_{ii}=0$, where 
$r_{ij}$ is the distance between sites $i$ and $j$. 

\noindent
The position of the $N^{th}$ conductance peak is given by the 
first difference of the ground state energy, namely, the 
chemical potential of the (isolated) dot, 
$\mu \equiv E(N+1)-E(N)$. The spacing between peaks is 
determined by the second difference (the inverse compressibility),   
\begin{eqnarray}
\chi_{N} \equiv E(N+1)-2 E(N)+E(N-1).
\label{eq_compress}
\end{eqnarray}
The occurrence of close peaks 
for certain values of electron number $N$ corresponds
to small values of $\chi_{N}$ (recall that
for a single island quantum dot, in which the single 
particle energies are neglected,
the inverse compressibility is a constant 
$1/C_{11}$). For the more general model described above 
the spacing distribution will of course fluctuate. 
In general, some constants appearing in the 
Hamiltonian $H=H_{c}+H_{sp}$ are 
random (e.g. the elements of the matrix $C$ and  the 
single particle energies $\epsilon_{i \alpha}$), 
but most experiments are performed on a single
quantum dot, so that fluctuations are meant with 
respect to the electron number $N$.
The numbers  $\chi (N)$ might then be
considered as values assumed by a random variable $\chi$
which has a certain distribution 
function $P(\chi)$. \\ 
\noindent
It might be instructive to compute the distribution $P(\chi)$
for a particular special case (albeit not realistic)
which looks deceptively simple. The reader 
who is not interested may skip the rest of this section. Take
$K>1$, $C_{ij}=0$ for $i \ne j$,  and $\epsilon_{i \alpha}=0$. 
The Hamiltonian $H=H_c$ is the classical energy of a set of 
$K$ charged metallic bodies which are very far apart from 
each other, so there is no coupling among them. 
This problem then belongs to a class 
of problems dealing with statistics of spectra 
of independent quantum systems \cite{Berry}. Here it refers to 
the addition spectra for a system composed of 
several independent subsystems, each of which has a certain 
spectrum which is a quadratic function of the number of 
particles residing on it. Thus, for example, 
by replacing $C^{-1}_{ii} \rightarrow \hbar \omega_{i}$,
it can represent the energy of $N$ spinless fermions distributed 
among $K$ independent oscillators.\\
\noindent 
Our aim is 
to find the distribution 
function $P(\chi)$. If charge is 
not quantized, the problem of finding the ground state 
energy is simply formulated as 
follows: Given $K$ independent capacitors $C_{i},\ i=1,\ldots,K$,
(in general, with random capacitances $C_{i}$), how                     
should the corresponding charges $Q_{i}$ 
be chosen so as to minimize the electrostatic energy
$E(Q)=\sum_{i=1}^{K} \frac {Q_{i}^2} {2 C_{i}}$
while holding the total charge 
$Q=\sum_{i=1}^{K} Q_{i}$ fixed?
This constrained minimum problem 
is trivially solved by requiring $\partial E(Q) /\partial Q =0$ 
and expressing $Q_{1}=Q-\sum_{i=2}^{K} Q_{i}$.
It is evident 
from dimensional arguments that $E(Q)$ is proportional 
to $Q^2$, with some coefficient denoted by $1/2C$, in which 
$C$ may be interpreted as the total capacitance of the system. 
Evidently, in this case the second derivative of $E(Q)$ with 
respect to $Q$ is independent of Q (and is easily calculated),
namely $\partial^2 E /\partial Q^2 = 1/C$.

\noindent
What happens then if charge {\em is} quantized? The formulation 
of the problem is now repeated albeit with 
$Q=N, Q_{i}=N_{i}$ (with $N$ and $N_{i}$ integers). 
Intuitively, one would 
expect $P(\chi)$ to be centered about $1/C$, but, surprisingly, 
this is not the case, and $P(\chi)$ is large near $\chi=0$. 
An analytic expression for $P(\chi)$ has been obtained 
recently by the authors \cite{JphysA}. 
In order to present it, let us 
assume that
the $K$ capacities $C_{i},\ i=1,2,\ldots,K,$ are rearranged so
that $C_{1}>C_{2}>C_{3}...>C_{K}$. It is also useful to 
divide all the capacitors by the largest one, to obtain
scaled capacitors $c_{i} \equiv C_{i}/C_{1}$. with
$1=c_{1}>c_{2}>c_{3}>\ldots>c_{K}$. 
Now let us define the following quantities:
\begin{equation}
a_{\ell} = \sum_{2 \le i_{1}<i_{2}<..<i_{\ell} \le K}
c_{i_{1}} c_{i_{2}} ...c_{i_{\ell}} \prod_{j \ne i_{m}}
(1-c_{j}), \qquad\ell=1,2,\ldots,K-1,
\label{eq_al}
\end{equation}
\begin{equation}
{\cal C} \equiv \sum_{i=1}^{K} c_{i},
\label{eq_C}
\end{equation}
\begin{equation}
B \equiv \prod_{i=2}^{K} (1-c_{i}).
\label{eq_B}
\end{equation}
With these notations we can provide a precise formula for
the spacing distribution. The formula relates to the density
function of the distribution and holds in the generic case (meaning that
for a specific set of measure 0 of parameters it is incorrect). The
following is obtained if the spacings are normalized to lie in the interval
$[0,1]$ (multiplying all of them by $C_1$):
\begin{eqnarray}
P(\chi)=\frac {1} {2 {\cal C}} \sum_{\ell=2}^{K} 
a_{\ell} \ell (\ell+1) 
(1- \chi)^{\ell-1} + \frac {B} {{\cal C}} \delta(1- \chi).
\label{eq_Pchi}
\end{eqnarray}
A plot of $P(\chi)$ is given in figure ~\ref{fig1}
%\begin{figure}[htb]
%\includegraphics[scale=0.75]{f1.ps}
%\caption{}
%\end{figure}
 and 
shows indeed that it has a maximum at $\chi=0$ and a 
delta function component at the inverse of the largest 
capacitance. In fact, 
when $K \rightarrow \infty$, the weight of the 
delta function shrinks to zero and (after an appropriate
renormalization) $P(\chi)$ approaches the 
Poisson distribution $e^{- \chi}$. \\ 
\noindent
\section{Results}
We now return to the full Hamiltonian 
of Eq. \ref{eq_H}. It contains 
the coefficients of capacitance and induction
matrix $C_{ij}$ and the single particle energies 
$\epsilon_{i \alpha}$ as input. In choosing 
the actual numerical values we use a few 
guidelines, one of which is to avoid too many
independent input data. As will be clear below, these are 
{\em not} fitting parameters, but rather a set of constants 
which are chosen once and for all on general physical grounds.  
\noindent
First, for the electrostatic part, recall that
the matrix $C$ should be a symmetric positive definite matrix 
with $C_{ii}>0$ and $C_{ij}<0$ for $i \ne j$. Besides, we 
expect it to be random. We then assume that $C_{ii}$ are 
random numbers uniformly distributed between $0$ and $W$,
whereas the nondiagonal elements are uniformly distributed 
between $-w$ and $0$. 
Actually, it is only the ratio $w/W$ which 
matters, so one may assume $W=1$ and use $e^{2}/W=1$ as 
an energy unit, leaving $w$ as a single constant reflecting the 
strength of coupling between islands. Practically, 
for each realization of $c_{ij}$'s drawn at random we check
whether the resulting matrix is positive definite, and reject
realizations which are not. Thus, we 
must put the restriction $w<1$ in order to generate a positive 
definite matrix $C$. Indeed, for $w>0.25$ (and $K=5$) most of the matrices 
generated randomly failed to be positive definite.
\noindent
Second, for the single particle energies,
we consider each electron 
island $i$ as a two-dimensional potential well 
$V_{i}(r)=\frac {1} {2} M \omega_{i}^{2} r^{2}$, where 
$M$ is an effective mass. 
One may then regard $\omega_{i}^{-1/2}$ 
as a measure of the radius of the corresponding 
electron island. Since the capacitance $C_{ii}$ 
is also proportional to this radius we assume 
$\omega_{i}= \gamma  C_{ii}^{-2}$, where $\gamma$ is 
a constant reflecting the relation between 
charging energies and single particle energies. The single particle 
energies in each electron island (subject to a perpendicular 
magnetic field $B$)
are then given explicitly. To be more specific, recall that
in two dimensions there are two quantum numbers for 
the orbital motion (denoted hereafter by $n,m$) to which 
we add a spin index $\sigma = \pm 1$. Then, 
with $\alpha =(n,m,\sigma)$, we have
\begin{eqnarray}
\epsilon_{i \alpha}= \frac {\hbar} {2} \left[ n \omega_{c} 
+ m \sqrt {\omega_{i}^{2}+\frac {1} {4} \omega_{c}^{2}}\ \right] 
+ g \mu_{B} \sigma B,
\label{eq_eps} 
\end{eqnarray}
where $\omega_c=eB/Mc$ is the cyclotron frequency, $g$ is the 
g-factor and $\mu_B$ is the Bohr magneton. Since $g$ contains 
the effective mass it is not known accurately. 
Its value is constrained on physical grounds (see below). 
Note that in this scheme the spacings between single particle 
energies are deterministic, and do not 
follow the Wigner surmise. The main cause of 
fluctuation is then due to the combination of non-random 
single particle energies and the occurrence of
numerous charging energies. Finally, the 
number of electron islands $K$ 
is determined by the size of the quantum dot.
The four input data of the model are then $K,w,\gamma$ 
and $g$.
Note that the gate voltage $V_g$ does not have an important role 
here. Indeed,
in actual experiments the variation of gate voltage serves to 
adjust the energies $E(N)$ with the chemical potential of 
the leads, but here the ground state energies are calculated directly. \\
\noindent
In order to avoid redundancy we are content with 
having a single set of these constants which is physically 
reasonable. In particular, it assures that the charging 
energy is much larger than single particle level spacings 
and that the Zeeman splitting is small at moderate magnetic fields.
Specifically, for $B=0$ we test the
distributions for $w=0.03,0.08$ and $0.20$, 
while for $B \ne 0$ we fix $w=0.03$. Besides, 
we take $\gamma=0.1$, and $g=0.0081$
 is chosen such that for moderate fields the Zeeman 
splitting is of the order of the mean level spacing.
Finally, the number of electron islands 
is fixed as $K=5$. Indeed, we use just a brute force trial algorithm, 
and hence cannot treat systems with a large number of electron 
islands. \\
\noindent
With these prescriptions
the ground state energies $E(N)$ and their
first and second differences are calculated 
for $B=0$ up to $N=200$ and for $B \ne 0$ 
up to $N=50$. 

\noindent
The first question we addressed is 
how the electrons are added among the islands. 
Figure~\ref{fig2} shows electron numbers $N_{i}$ in each
of the five islands, {\em v.s} the total electron number $N$, in 
the range $60<N<80$ 
(for $w=0.03$). Evidently, the order of curves is according 
to the value of the capacitance $C_{ii}$. On a larger scale, the 
numbers $N_{i}$ grow linearly with $N$, as it should be. Let us
then consider the question of redistribution. From figure~\ref{fig2} 
we see that redistribution occurs only once, as $N$ grows between 
$67$ and $68$ (see the vertical line). The addition of an electron
to the second (or third) island involves also a transfer of an 
electron from the fourth island to the second (or the third) one. 
This scenario occurs also in other ranges of $N$ with the same 
proportion (namely about four percent). Thus, within the present model, 
redistribution is present, although it is rare and minimal.\\
\noindent
The next question we address is how the 
ratio between the inductive and capacitive 
coefficients (represented here by the parameter $w$) 
affects the distribution $P(\chi)$. 
It was shown 
in the previous section that for $w=0$ the 
distribution has a maximum at $\chi=0$. It is 
intuitively expected (based on the concept of level 
repulsion) that for $w \ne 0$ the maximum will be shifted 
away from zero. This is also verified by our 
numerical results but, somewhat unexpectedly, 
the peak at small $\chi$ persists
even at $w=0.20$. The distribution 
$P(\chi)$ is drawn in figure~\ref{fig3} for 
$w=0.03, 0.08$ and $w/W=0.20$. The normalization is such that 
the largest capacitor is $C_{11}=1.0$, so that if 
other capacitors are absent the distribution would have 
a delta function at $\chi=1$. 
In all three cases 
the peak position is 
much smaller than $1$. 
 Recall,
however, that besides the capacitance induction matrix $C$ the 
total energy is determined also by the 
single particle energies in each island. The remarkable 
point is that the distribution is not Gaussian (which is 
the hallmark of spacing distributions in small quantum dots).\\
\noindent
Finally, we check the behavior of the first difference 
$E(N+1)-E(N)$ as a function of the magnetic field. Recall that 
this quantity is proportional to 
the position of the $N^{th}$ Coulomb blockade 
peak. As a measure of the strength of the magnetic field we 
use the parameter $\omega_{c} / \omega_{0}$, where $\omega_{0}$ 
is the harmonic oscillator frequency of the largest island. 
The positions of the peaks for $39<N<48$ 
and $w/W=0.03$ are displayed in 
figure~\ref{fig4}. A comparison with the results 
displayed in figure 2 of Ref. \cite{Ashoori3} indicates
a remarkable qualitative agreement. In particular, 
it shows that groups of two (and sometimes even three) 
electrons can tunnel through the quantum dot at almost 
the same gate voltage. The oscillations at small 
magnetic field just mark transitions to lower Landau levels 
as the magnetic field increases. The phenomena of alternate 
bunching $(N,N+1) \rightarrow (N-1,N)$ is also reproduced 
in the present picture. \\
\noindent
In conclusion, we suggest a classical 
model in which a large  semiconductor quantum 
dot is viewed as a collection of metallic 
electron islands with capacitive and inductive 
coupling among them. The effect of adding a magnetic field 
is manifested through its orbital as well as its 
spin contents. The model 
can explain the occasional occurrence of 
couples or even triples of closely spaced Coulomb blockade peaks, 
as well as the qualitative behavior of peak positions 
with the magnetic field. 
The results of the previous section displayed in 
figure~\ref{fig1} together with 
an analysis of the results displayed in figure~\ref{fig3} 
can provide an answer to the question why, in the present 
model, there are couples and sometimes even triples of 
close Coulomb blockade peaks. When the capacitors are 
completely independent, the distribution is close to 
Poissonian as can be proved analytically. Introducing 
an inductive coupling spoils this picture, but does 
not destroy it completely, so that even at $w=0.2$
there is still a probability for the occurrence of 
close peaks. 
The eventual decrease of the distribution 
near $\chi=0$ seems to be related to the fact that, unlike 
the case of independent systems, the capacitance matrix 
is non-diagonal. Switching on the coupling here then  
has an effect similar to a weak ``spacing repulsion'', 
similar to the  familiar effect of perturbation in
 two level systems.
\begin{acknowledgments}
This research was supported in part by 
a grant from the Israel Academy of 
Science and Humanities under {\em Centers of Excellence Program}.  
One of us (Y. A) is grateful to 
R. Ashoori for discussion and suggestions.
\end{acknowledgments}

\newpage
\section{Figure Captions}
\begin{figure}
\caption{Distribution of inverse compressibility $P(\chi)$ 
for a system containing independent subsystems each of 
which has a ground state energy proportional to the 
square of the number of particles it contains. The 
graph corresponds to a special case of equation 
\ref{eq_Pchi} with $K=5$.
\label{fig1}}
\end{figure}

\begin{figure}
\caption{
Distribution of electrons among the five islands 
as a function of the total electron number $N$ at 
zero magnetic field. 
The choice of constants $K=5$, $w=0.03$ is
explained in the text.
The number $N_i$ of electrons on island $i$ is commensurate 
with its capacitance $C_{ii}$. In most cases the 
addition of an electron does not perturb the occupation 
of other islands. An example of redistribution is marked 
with a vertical line. An addition of an electron causes 
a minimal redistribution.
\label{fig2}
}
\end{figure}
%
%\begin{figure}
%\caption{
%Inverse compressibility $\chi_N$ as function of 
%electron number $N$ in the dot. The units on the ordinate 
%are the charging energies of the largest island, 
%namely, $e^{2}/W=1$ (see text).
%\label{fig3}
%}
%\end{figure}

\begin{figure}
\caption{Effect of strength of inductive coupling $w$ on the
distribution $P(\chi)$ of level spacings in the dot 
at zero magnetic field (not normalized). It is 
displayed for $K=5$ and $w=0.03, 0.08$ and $0.20$ in (a), (b) 
and (c) respectively. 
\label{fig3}} 
\end{figure}

\begin{figure}
\caption{Coulomb blockade peak position for electron number 
$N$ between $39$ and $48$ as functions of the magnetic field. 
Here $\omega_{o}$ is the oscillator frequency of the largest 
island and $\omega_c$ is the cyclotron frequency. The units on 
the ordinate are energy units as explained previously. They 
are proportional to the gate voltage appropriate 
for the corresponding peak. The values chosen for 
the constants $w=0.03$, $K=5$, $\gamma=0.1$ 
and $g=0.0081$
are explained in the text.
\label{fig4}}
\end{figure}

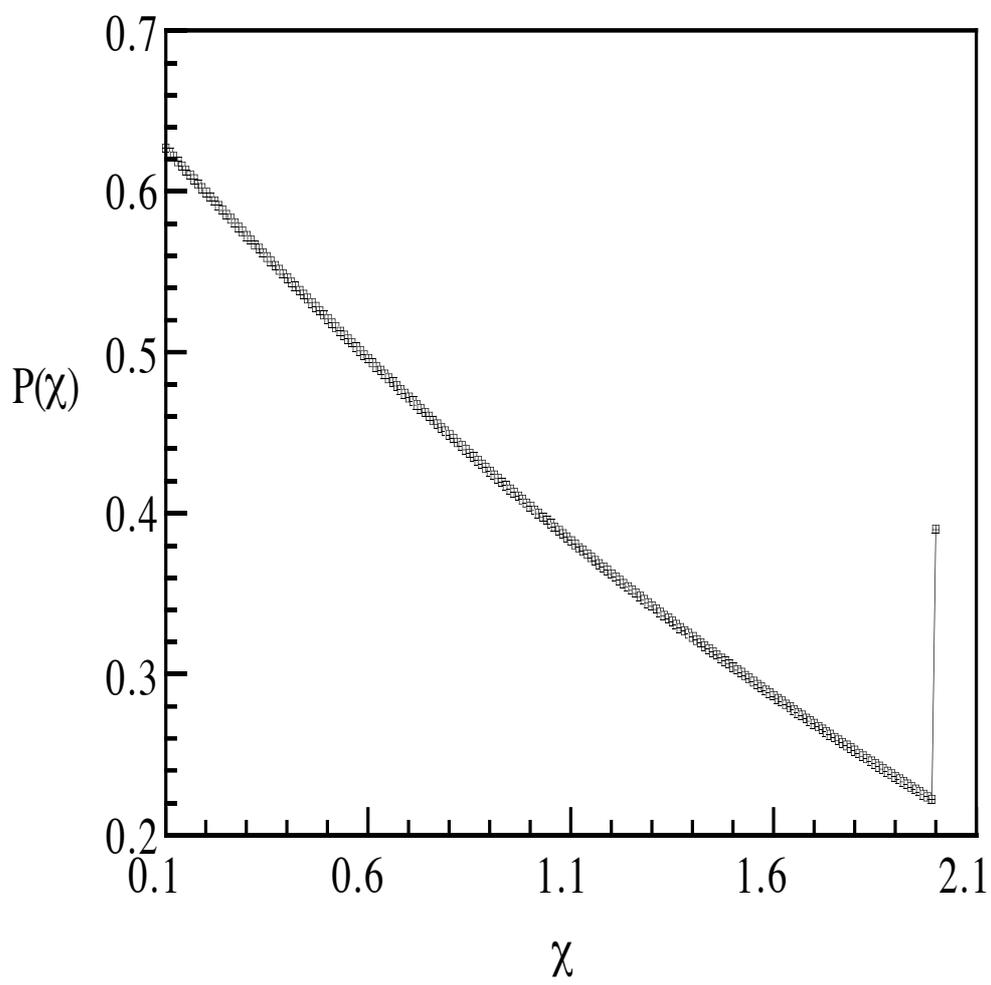
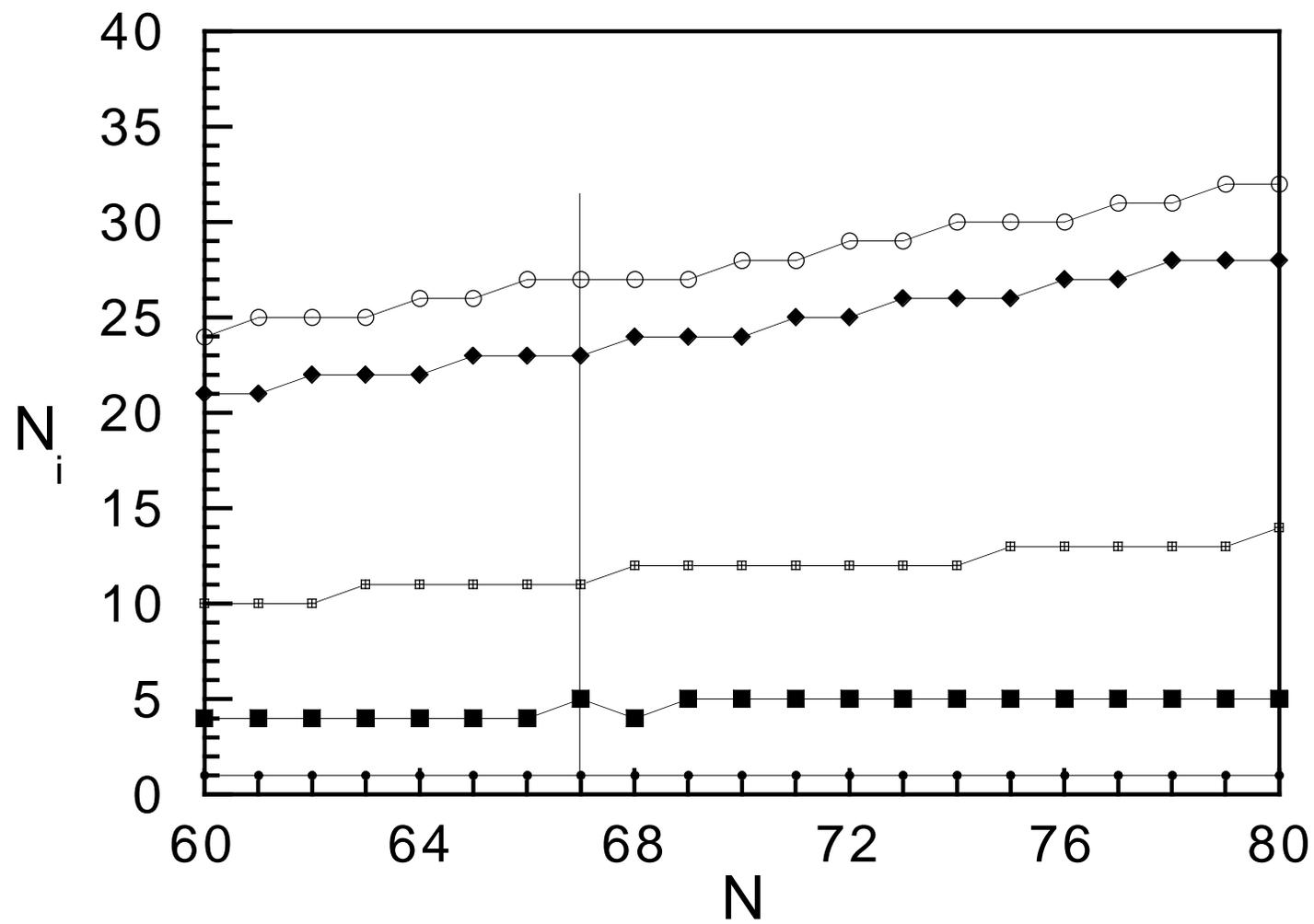
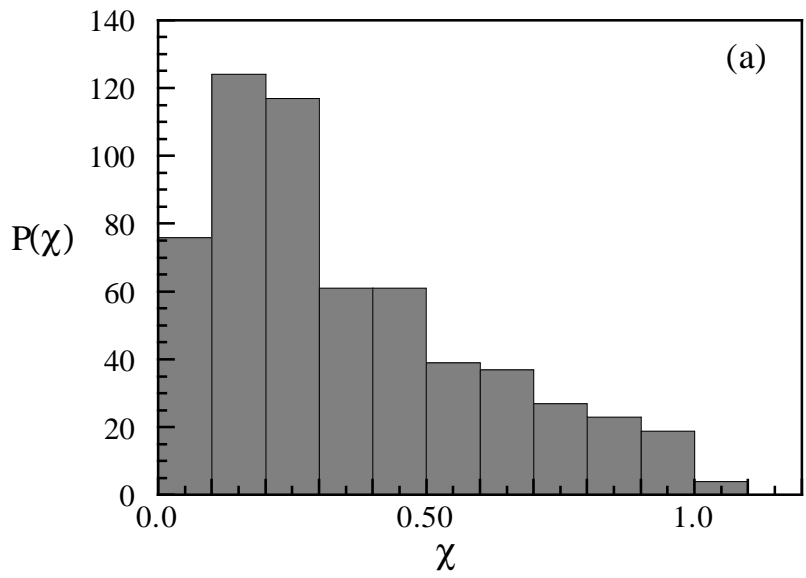
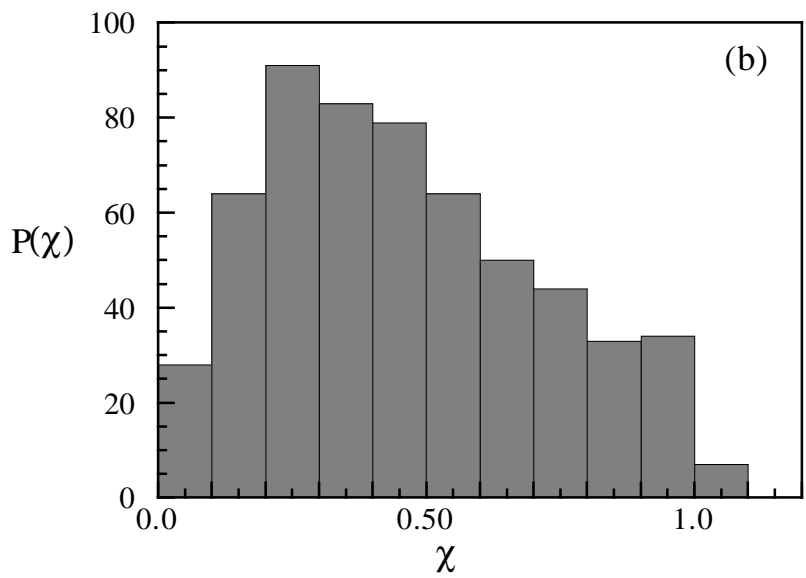
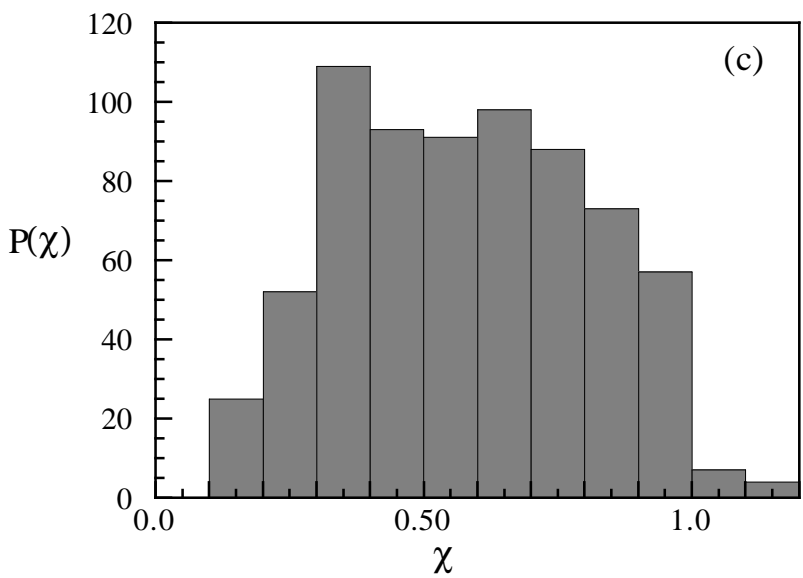
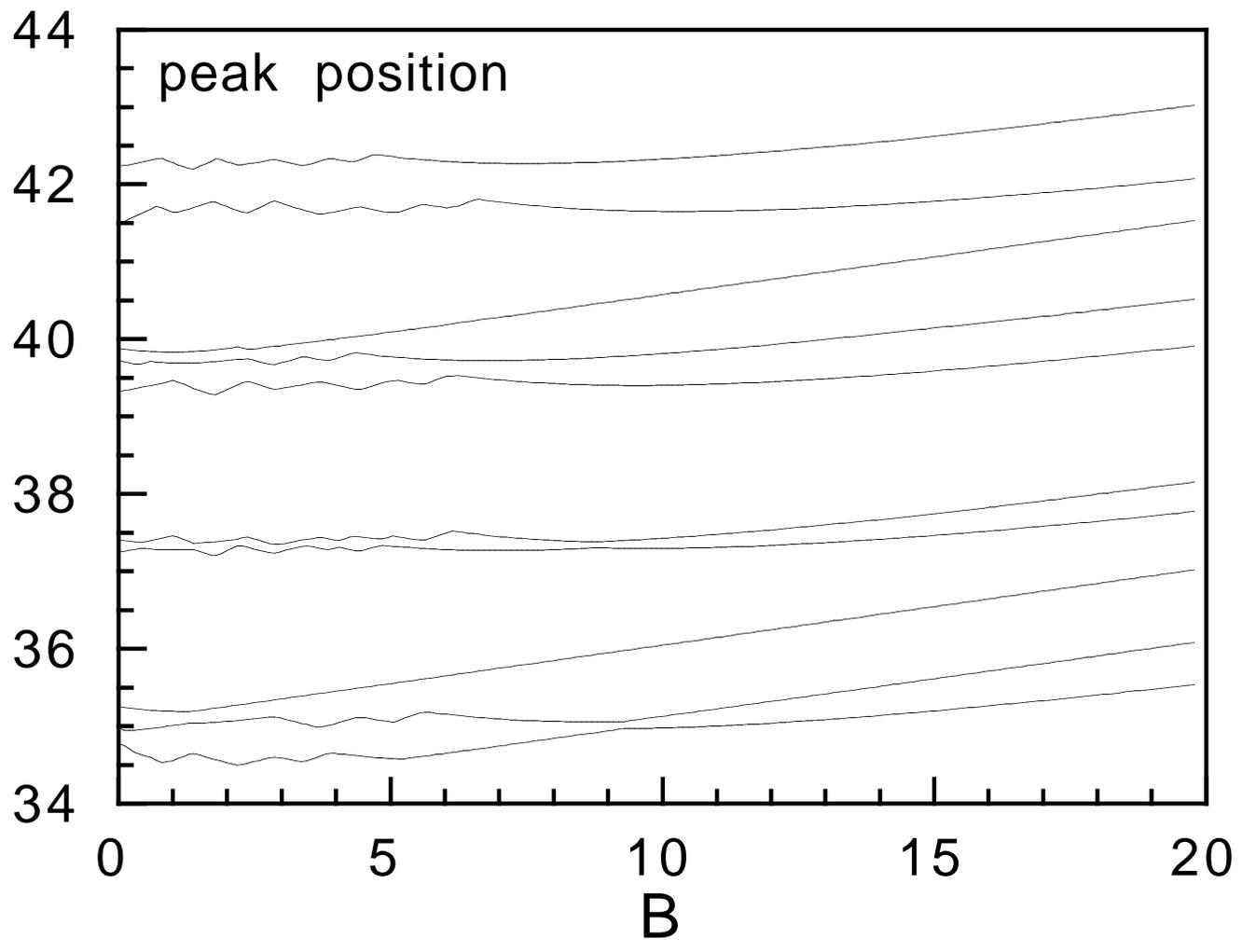
\end{document}